\documentclass[graybox]{svmult}

\usepackage{amssymb} 
  
\usepackage{mathptmx}       
\usepackage{helvet}         
\usepackage{courier}        
\usepackage{type1cm}        
\usepackage{makeidx}         
\usepackage{graphicx}        
\usepackage{multicol}        
\usepackage[bottom]{footmisc}

\def\0#1{{\mathrm{#1}}}
\def\1#1{{\mathbb{#1}}}
\def\2#1{{\mathbf{#1}}}  
\def\3#1{{\mathcal {#1}}}
\def\4#1{{\mathsf {#1}}}   
\def\5#1{{\widetilde{#1}}} 
\def\6#1{{\overline{#1}}} 
\def\7#1{\breve{#1}}
 \def\8#1{{\widehat{#1}}}  
 \def\9#1{{\check{#1}}}
 \def\fr#1{{\mathfrak{#1}}}

\def\io{{\sc io}}

\def\miss{\mathrel{{\kern3pt\backslash\kern-8.1pt\bigcirc}}}

\def\O+{\bigoplus}
\def\LA+{\;\circ\hskip -9.1pt +\,}
\def\nogo{\mathrel{+\kern-9.0pt + \kern-12.4pt\to}}

\def\v{{\vee}}

\def\Bar{{{}\vrule height 8pt width 1pt depth 3pt{\kern3pt}}}
\def\<{{\left<\right.}}
\def\>{{\left.\right>}}

\def\Cliff{\mathop{{\0{Cliff}}}\nolimits}
\def\Grade{\mathop{{\mathrm{Grade}}}\nolimits} 
\def\Grass{\mathop{{\mathrm{Grass}}}\nolimits} 
 
\def\Dim{\mathop{{\mathrm{Dim}}}\nolimits} 

\def\Dual{\mathop{{\mathrm {Dual}}}\nolimits}
\def\Endo{\mathop{{\mathrm {Endo}}}\nolimits} 

\def\hexp{\mathop{{\mathrm{hexp}}}\nolimits} 

\def\Qi{{\8{\imath}}}

\def\so{\mathop{{\mathrm {so}}}\nolimits}
\def\SO{\mathop{{\mathrm {SO}}}\nolimits} 
\def\slin{\mathop{{\mathrm {sl}}}\nolimits}

\def\spin{\mathop{{\mathrm {spin}}}\nolimits}
\def\su{\mathop{{\mathrm {su}}}\nolimits} 

\def\Z{{
\mbox{\scriptsize 1}
}}

\def\v{\vee}
\def\ox{\otimes}

\def\x{\times}

\def\BEQ{\begin{equation}}
\def\EEQ{\end{equation}}
\def\BEN{\begin{enumerate}}
\def\EEN{\end{enumerate}}
\def\BI{\begin{itemize}}
\def\BEI{\end{itemize}}
\def\BEA{\begin{eqnarray}}
\def\EEA{\end{eqnarray}}
\def\BTA{\begin{table}}
\def\ETA{\end{table}}

\def\fro{\leftarrow}

\def\cto{\;\circ\kern-4pt\to\;}
\def\cfro{\;\fro \kern-4pt\circ\;}

\def\rar{\rightarrow}

\def\mapsfro{{\fro}\kern-4pt\rule{.5pt}{5pt}}

\def\dar{{\downarrow}}

\def\Diagr#1#2#3#4#5#6#7#8
{\begin{array}{rcccl}%
&              &{\scriptstyle #5}           &                                            &\cr
& #1                                          & \rar& #2                                      &\cr
\cr
{\scriptstyle #8}&\dar \;\; &         &\;\;\dar&{\scriptstyle#6} \cr
\cr
& #4                                   & \,\rar \,& #3                                      &\cr
&                           &{\scriptstyle#7}&  			             &     
            \end{array}}

\newtheorem{unitas}{}[section]

\title*{PALEV STATISTICS AND  THE CHRONON}
\author{David Ritz Finkelstein}
\institute{Georgia Institute of Technology, Atlanta, Georgia.
\email{finkelstein@gatech.edu}}
\begin{document}
\maketitle

\abstract
{A finite relativistic quantum space-time is constructed.
Its unit cell has Palev statistics defined by a spin representation of an orthogonal group.
When 
the Standard Model and general relativity are  physically regularized by such space-time quantization, 
their gauges are fixed by nature;  the cell groups remain.
}
 
\section{\em Novus ordo seclorum}

The goal is still a finite physical theory that fits our finite physical experiments.
 The classical space-time continuum led to singular (divergent) quantum field theories:
Infinity in, infinity out.
In ancient times, a continuum was the only way to understand
the translational and rotational invariance of Euclid's geometry.
Today there are  quantum spaces with a finite number of quantum points
that  still have the continuous symmetries of
gravity and the Standard Model,
at least within experimental error.
If electron spin were any simpler it would not exist.
The strategy is build the cosmos from such atoms.

Quantum spaces are represented by probability spaces that
define the lowest-order logic of their points.
Modular architecture requires the higher-order logic, classically  dealt with by set theory.
Classical space-time and field theory are formulated 
within classical set theory;
perhaps quantum space-time and field theory 
need  a quantum set theory.
Classical set theory was invented by Cantor to represent the mind of the Eternal.
Quantum set theory is intended to represent 
the system under study as quantum computer.
Like any quantum theory it statistically represents   input/outtake (\io) beams of systems
by what
Heisenberg called {\em probability vectors}\/.

Quantum logic is a square root of classical logic: 
Transition probabilities in the classical sense are squares of  components of the probability vector, transition probability amplitudes.
Call the quadratic space of probability vectors a probability space $\3P$.
In  general $\3P$, like the quantum space of Saller \cite{SALLER2006a},  includes both input (ket) and outtake (bra) vectors.
Distinguish these by the signs of their norms. 
$\3P$ is not a Hilbert space.

A quantum theory should describe 
populations as well as individuals.
Enrich the  quadratic probability space $\3P$ to a 
probability algebra $\3P$
whose product $a b$
represents successive application of input/outtake (\io) operators. 
The one-system probability vectors form
a generating subspace $\3P_1\subset \3P$.
$\3P$ consists of polynomials  over $\3P_1$,
subject to constraints and identifications
said to define the statistics.

A  regular quantum theory is one with a finite-dimensional 
probability algebra
\cite{BOPP1950}.

The probability algebra $\3P^-$ for fermions 
is a Clifford algebra, defined by anti-commutation relations
among the one-fermion vectors:
\BEQ
\forall x\in \3P^-_1:\quad x^2=\|x\| = x\cdot x\/.
\EEQ
Its dimension is $\Dim\3P^-=2^{\Dim \3P_1}$\/, 
so
Fermi statistics is regular if the one-fermion probability space is.

The probability algebra $\3P^+$ for even quanta is commonly  
assumed to be a Bose (Heisenberg, canonical) algebra whose generators obey
\BEQ
\forall x\in \3P_1:\quad xy-yx =\epsilon(x,y)
\EEQ
with a given skew-symmetric bilinear form $\epsilon$ 
on $\3P_1\ox\3P_1$\/.
This is not exactly right:
 Bose statistics is always singular.
 Pairs of fermions, however, obey a regular statistics
whose  probability algebra
envelops a Lie algebra,
 with Bose statistics and the Heisenberg algebra 
 as a singular limit.
 This is a special case of Palev statistics  \cite{PALEV1977, PALEV2002}, which is
reviewed next.

\section{Palev statistics}
\label{S:PALEV}

For any semisimple Lie algebra $\fr p$,
a{ \em Palev statistics of the $\fr p$  class}
 is one whose probability algebra $\3P$ 
 is a finite-dimensional enveloping algebra
 of $\fr p$
(has $\fr p$ as commutator Lie algebra).

Fermi and Bose statistics 
have  graded Lie algebras $\fr f,\fr b$ that specify their commutation relations.
$\fr f$ and $\fr b$ as vector spaces are also one-quantum probability spaces.
They have essentially unique  irreducible unitary representations;
these serve as many-quantum probability spaces.
There are, however, an infinity of  irreducible representations 
of a Palev algebra $\fr p$
that might serve as many-quantum probability space.
Empirical choices must be made for Palev statistics
that are already decided for Fermi and Bose statistics.

Palev gives a representation of $\slin(n+1)$ on a Fock space $W_p$ of symmetric tensors of degree $p$. 
$W_p$ is a Hilbert space, appropriate for his applications and not for
these. 

The  probability space of a hypothetical quantum event must  have enough dimensions to allow for the observed quantum systems.
It is not clear that events in space-time can be experimentally located to 
within much less
than a fermi,
corresponding to a localization in time of about $10^{-25}$ s.
The Planck limit at $10^{-43}$ s was initially a conjecture
based on pure quantum gravity.
Our instruments, to be sure, do not seem to be  made of gravitons alone, 
but take part in all the interactions.
The Planck time seems at best a poor lower bound to the quantum of time.

However the energy at which all the running coupling constants
seem to converge is not very much greater than
the Planck energy,
so it may indeed have universal significance.
Yet crystals have many scales besides cell size, such as
Debye shielding length, skin depth, coherence length, and mean free paths.
The Planck time and the unification energy  might correspond more closely to one of these
than to the cell size $\4X$.
To avoid a premature commitment,
call the natural time $\4X$ the {\em chrone}\/.

How many dimensions must the event probability space have? 
Suppose the lifetime of the four-dimensional universe is $10^{21}$ s; 
an error by a factor of 100 will not matter much.
If $\4X\sim T_{\0P}\approx 10^{-43}$ s 
then the dimensionality of the  
history probability space of the 
cosmos---which we cannot observe maximally---is 
about $10^{1024}$. 
The largest system that can be maximally 
observed by a co-system within such a cosmos---here 
we renounce the perspective of the Eternal---is much simpler.
Its probability space might have no more than
$\log_2 10^{256}\sim 3000$ dimensions.

Here are two examples of Palev statistics:

\subsection{Rotatons} The $\so(3)$ Lie algebra 
with commutation relations relation $\2L\x\2L=\2L$ 
defines an aggregate of palevons of the $\so(3)$ kind, 
whose quanta must be called rotatons,
since Landau preempted the term ``roton".
Relative to an arbitrary component $L_3$ as generator of a 
Cartan subalgebra,
 the root vectors $L_{\pm}=L_1\pm  i L_2$
represent the input and outtake of spin-1 rotatons.
An irreducible representation with extreme eigenvalues $\pm il$ for $L_3$
represents a Palev statistics with no more than $2l+1$ rotatons present in a single
aggregation.

\subsection{Di-fermions}  
\label{S:DIFERMION}

  Di-fermions are palevons. 
  If the fermion probability space is
  $2N\1R$ then
  the di-fermion  is a palevon of type $\so(N,N)$.

This refers to the elementary fact that the Fermi commutation relations 
for a fermion with $N$ independent probability vectors
define a Clifford algebra $\Cliff(N,N)$,
 and the  second grade of $\Cliff(N,N)$ is
 both the probability space for a
fermion pair, and
 the Lie algebra  $\spin(N,N)$ 
defining a Palev statistics of class $D_N$.

\subsection{Regular space-times}
\label{S:REGULAR}

Call a  spacetime regular if its coordinate algebra is regular.
All its coordinates then have finite spectra.
There are not many regular spacetimes in the literature.

Singular (non-semisimple) Lie algebras can be regularized
by slightly changing some 
vanishing commutators,
undoing the flattening contraction that led, presumably, from the regular to the singular.

The prototype of such regularization by decontraction is (special) relativization \cite{SEGAL1951,INONU1952}.
This regularizes the Galilean
Lie algebra $\fr g =\fr g(\2L,\2K)$
of Euclidean rotations $\2L$ and Galilean boosts $\2K$,  
to the Lorentz Lie algebra
$\so(3,1)$.  
Write such relations as
\BEQ
\so(3,1)
\cto
\fr g
\quad  
\mbox{or}\quad \fr g{\;\fro \kern-4pt\circ\;}\so(3,1),
\EEQ
directed from the regular algebra
to the singular.

The sole remaining culprit today is Bose statistics, 
the canonical Lie algebra.
Its de-contraction requires 
adding new variables.
This is the general case;
special relativity and quantum theory 
were exceptional in this respect.

Some  physical self-organization must then freeze these extra variables out
near  the singular limit.
This can be tested experimentally in principle by disrupting this organization.
Hopefully,  a suitable regularization of
 the remaining singular theories
 will once again improve the  fit with experiment.

The Killing form of classical observables  is as singular as can be:
identically 0.
It is nearly regularized 
by canonical quantization:
\BEQ
a_{\0{comm}}({\2x},{\2p},i) \cfro\; \fr h(N) \/,
 \EEQ
 where  the canonical (Heisenberg)
 Lie algebra is
 \BEQ
  \fr h(N): [x^{\nu},p_{\nu'}]=i\hbar \delta^{\nu'}_{\nu},\quad 
  \nu,\nu' = 1,\dots, N,
  \EEQ
  other commutators vanishing.
The solvable radical  $\1C$ generated by $i$ survives 
canonical quantization.
Recall  that $\fr{hp}$ 
does not fit into any $\slin(N\1R)$.
(The left-hand sides of the canonical commutation relations 
would have well-defined trace 0, 
and the right-hand side $i$ would have non-zero trace.)
Canonical quantization is a quantization interrupted
by premature canonization.

\subsection{Feynman space-time}  
\label{S:FEYNMAN}

Feynman  \cite{FEYNMAN1941} seems to have constructed the first
regular relativistic quantum space-time $\3F$\/.
Its positional coordinates  are finite spin sums:
\BEQ
\label{E:FEYNMAN}
x^{\mu}\cfro\; \;\8x^{\mu}=\4X\left[ \gamma^{\mu}(1)+\dots +\gamma^{\mu}(N) \right],\quad \mu=1,2,3,4.
\EEQ
The $\gamma^{\mu}$ have unit magnitudes, and
$\4X$ is the natural quantum unit of time, or {\em chrone}.
If the commutators $[\gamma^{\mu'}(n'), \gamma^{\mu}(n)]$
vanish for $n\ne n'$ then the probability vector space for $\3F$
is a $16^N$-dimensional Clifford algebra.
Quantification theory abbreviates (\ref{E:FEYNMAN}
to
\BEQ
\8x^{\mu}=\4X \6{\psi}\gamma^{\mu}\psi.
\EEQ

Each term in the sum represents a hypothetical 
quantum element of the space-time event;
call it a chronon.
The Feynman chronon has spin 0 or 1, because 
$\gamma^{\mu}$ 
has both a scalar part $\gamma^0$ 
and a vector part $(\gamma^1, \gamma^2, \gamma^3)$.

Nature seems to have a unit of space-time size like $\4X$ at every event,
fixing the gauge in the original sense of Weyl.
If each event has a space-time measure $\4X^4$, then
the dimension  of the one-event probability space is proportional to the space-time volume;
as if the event statistics is extensive in the sense of
Haldane \cite{HALDANE1991, PALEV2002}.

\subsection{Yang space-time}

Yang \cite{YANG1947} proposed the first regular 
relativistic space-time-momentum-energy  Lie algebra,
leaving the signature somewhat open:
\BEQ
\fr y =(\so(5,1) \;\mbox{or}\; \so(3,3))\;\cto \fr{hp}(4) .
\EEQ
$\fr y$ can represent the orbital variables 
of a spinless relativistic quantum.
 Yang restricted consideration to representations in Hilbert space,
  however,
blocking regularity.
Regular version of the Yang theory uses a finite-dimensional representation
of the Yang Lie algebra $\so(3,3)$.
Its probability algebra must then have  an indefinite norm (\S\ref{S:INDEFINITE}). 

The Yang Lie algebra $\fr y$
is not to be confused with  the conformal $\so(3,3)$ Lie algebra.
They are isomorphic but act on different physical variables
and have different physical effects.
We deal with groups of physical operations, not abstract groups.

\subsection{Interpretation of the indefinite norm}
\label{S:INDEFINITE}

In special relativity the sign of the Minkowski metric form $\4g$ distinguishes  allowed (timelike) directions
from forbidden (space-like) ones.

In a relativistic quantum theory of the Dirac kind,
the probability-amplitude form $\beta$
is neutral (of signature 0).
\BEQ
\beta\Psi\Psi=\Psi^{\beta}\circ \Psi
\EEQ
 gives  the mean flux of systems from the experiment (not the absolute flux).
The sign 
 distinguishes input probability vectors  from outtake probability vectors
\cite{DIRAC1974}.

Let source kets bras have positive norm and sink bras negative.
This changes no physics in the usual quantum theory,
which does not add bras and kets.
Here it enlarges the group  and must be tested by experiment.

A regular theory might  associate an elementary particle 
with an irreducible  finite-dimensional isometric representation
of a simple Lie algebra $\fr y$ that approximates the Poincar\'e Lie algebra:
\BEQ
\label{E:PARTICLE}
\fr y\;\cto \fr{hp}\/.
\EEQ
Then all one-particle observables have finite spectra. 

The  constant $i\hbar$ of the usual quantum physics 
is then another non-zero vacuum expectation value,
a frozen variable like the Minkowski metric $g_{\mu'\mu}$
and the Higgs field.
Centralizing a hypercomplex number by a condensation
gives mass to any gauge boson that transports that number;
this was shown for the quaternion case, for example.
Such a frozen $i$ must be assumed in the  Yang and Segal
space-time quantizations \cite{YANG1947, SEGAL1951}
based on the Lie algebra
\BEQ
\label{E:YANGLIE}
\fr y = \so(3,3) \cong\spin(3,3)\cong \slin(4\1R) .
\EEQ

Infinitesimal generators of
$\so(3,3)\cong \spin(3,3)\cong \slin(4\1R)$  make up a tensor 
$[L_{y'y}]$ ($y,y'=1,\dots, 6$)
with 15 independent components,
representing orbital variables of the Yang scalar quantum.
This $\fr y$ is also a candidate for the scalar particle Lie algebra
$\fr y$ of (\ref{E:PARTICLE}).
The Feynman quantum space-time and the Penrose quantum space 
\cite{PENROSE1971} can be regarded as 
spin representations of the Yang Lie algebra.

$\Qi$, the quantized $i$, is a normalized $2\x 2$ sector 
$[L_{z'z}]$ 
($z,z'=5,6$) of $[L_{y'y}]$
 in an adapted frame.
 The   tensor $[L_{y'y}]$ then breaks up according to
\BEQ
[L_{y'y}] = \left[\begin{array}{l|ll}
L_{\mu'\mu}&ix_{\mu'}&ip_{\mu'}\\
\hline
ix_{\mu} &0&L_{56}\\
ip_{\mu} &L_{65}&0
\end{array}
\right]\sim \left[\begin{array}{l|l}
4\x 4&4\x2\\
\hline
2\x 4&2\x 2
\end{array}
\right]\/,\quad y,y'=1,\dots, 6,
\EEQ
which includes the Lorentz generator $L_{\mu'\mu}$ as a $4\x 4$ block,
position $x^{\mu}$ and momentum $p^{\mu}$ ($\mu,\mu'=1,2,3,4$)
as $4\x 1$ blocks, and $L_{z'z}$ ($z,z'=1,2$) as a $2\x 2$ block.

Posit a self-organization, akin to ferromagnetism, that causes
the extra component $L_{65}$
 to assume its  maximum magnitude
in the vacuum.
Small first-order departures from perfect organization of $i$ 
make second-order errors in $|i|$.

\subsection{Locality}

One more limit stands between the regular Yang Lie algebra
$\fr y$
and singular canonical field theories.
The special-relativistic kinematics and $\fr y$
have a canonical
symmetry  between the $x^{\mu}$ and $p^{\mu}$.
Yet there are great physical differences 
 between these variables.
Under the
composition of systems,  $p_{\mu}$ is extensive
and $x^{\mu}$ is intensive.
The fundamental gauge interactions of the Standard Model and gravity are local in $x^{\mu}$
and not in $p_{\mu}$;
unless asymptotic freedom
can be regarded as a weak form of locality
in $p_{\mu}$.

This suggests that there is a richer  class of regular quantum structures 
that have classical differential geometry and gauge field theories
as organized singular limits,
with at least three quantification levels:
the chronon, the event, and the field.

Wigner proposed that an elementary particle 
corresponds to an irreducible unitary representation 
of the Poincar\'e group.
Up-dates in this concept are called for by his later work.
The Wigner concept of elementary particle  
gives no information about  interactions between particles.
Gauge theory requires an elementary particle
 to have a location in space-time
where it interacts
with a gaugeon.
It would then seem useful to associate 
an elementary particle with  an irreducible representation 
of the Heisenberg-Poincar\'e Lie algebra 
$\fr{hp}(x^{\mu}, p_{\mu}, L_{\mu'\mu}, i)$ instead,
which fuses the Poincar\'e 
and the Heisenberg (canonical) Lie algebras.
This is the Lie algebra that Yang regularized.

\subsection{Quantization and quantification}

Quantification and 
quantization 
are related  like  archeology 
and architecture.
They concern similar structures,
but quantification synthesizes them from the bottom up,
which likely the order of formation, 
while quantization analyzes them from the top down, the order of discovery.

Quantization re-introduces a quantum constant
that  the classical limit eliminates.
Quantification does not introduce one because  the quantum individual
already provides it.

In particular, space-time quantization 
introduces a new quantum entity, the chronon, carrying 
a time unit, the chrone  $\4X$,  and an energy unit, the erge  $\4E$.
The chronon is no  particle in the usual sense 
but a least part of the history 
of a particle. 

Canonical quantization can also be interpreted as a  quantification with a singular statistics.
What is sometimes called ``second quantization" is more accurately 
a second quantification.

\subsection{Gauge}

A gauge is an arbitrarily fixed  movable standard used in measurements.
It is
part of the co-system, the complement of the system in the cosmos.
As part of the co-system, a gauge is normally studied under low resolution and
treated classically.
A field theory may postulate
 a replica of the gauge at every event in space-time,
 forming an infinite  field of infinitesimal gauges.
 Weyl's original
gauge was an infinitesimal  
analogue of a carpenter's gauge or a machinist's gauge block,
a movable standard of length; 
hence the name.

A gauge transformation changes the gauges
but fix the system.
They form a Lie group, 
the gauge group,
which indicates the arbitrariness of the gauges.
t is customary to assume that the relevant dimensions of the gauge field are fixed during an experimental run, 
so that the experimental results can be compared meaningfully 
 with each other.
This means that the gauge 
must be  stiff.
For example,  machinist's gauge blocks are often made of tungsten carbide.
Such rigid constraints become a source of infinities.

In a simple quantum theory, however, all variables have discrete spectra.
All eigenvalues can be defined by counting instead of by measuring.
There is no need for arbitrary units, external gauges, or gauge group;
Nature provides the gauge within the system.
Thus a gauge group is another sign of interrupted quantization.

One well-known way to break a gauge group is by self-organization.
It is often supposed that the Higgs field, which breaks an isospin group,
is such a condensate.

A gauge group is also broken, however, 
when further quantization discovers a natural quantum unit, 
fixing a gauge.
The quantized $i$ that breaks $\su(2)$ and imparts mass in quaternion quantum gauge theory is of that kind.
So is the quantized imaginary $\Qi$ that breaks Yang $\so(3,3)$.
 The Higgs $\8{\eta}$ that breaks isospin $\so(3)$,
and the gravitonic $\8{g}_{\mu'\mu}$ that breaks $\slin(4\1R)\cong 
\so(3,3)$  may also be such natural quantum gauges,
to be recovered by  regularizing the kinematical Lie algebra of the Standard Model through further quantization.

Notation: The one-quantum total momentum-energy vector is, up to a constant, the differentiator $[\partial_{\mu}]$, canonically conjugate to the space-time position vector $[x^{\mu}]$.
$[\partial_{\mu}]$ reduces to a gauge-invariant differentiator $[D_{\mu}]$, also canonically conjugate to $x^{\mu}$, related to kinetic energy,
 and
a vector $\Gamma_{\mu}(x)$  that commutes with position, related to potential energy:
\BEA
\label{E:TOTAL}
\partial_{\mu}&=&D_{\mu}\kern 18pt+\kern 15 pt\Gamma_{\mu}. \cr
\0{Total} &=& \0{Kinetic }+\0{ Potential}
\EEA

The gauge  commutator algebra $a(x^{\mu},
D_{\mu}(x), F_{\mu'\mu}, \dots )$ is generated by the space-time coordinates $x^{\mu}$  
and the kinetic differentiator $D_{\mu}(x)$, and includes the 
field variables $F_{\mu'\mu}=[D_{\mu'},D_{\mu}]$ and their higher covariant derivatives.
Its radical  includes all functions of the $x^{\mu}$.  
This makes it singular too.

Gauging semi-quantizes.  It converts  the commutative
operators $\partial_{\mu}$ into  the non-commutative ones $D_{\mu}$,
and for  individual quanta these are observables.
Its  contraction parameter is the coupling constant.
Landau quantization in a magnetic field is of that kind.

Gauging also quantifies:
It converts  one finite-dimensional global gauge group
$G$ into many  isomorphs of $G$, 
one at each space-time point. 

Gauging introduces infinities because the number of gauges
is assumed to be infinite.
Thus quantum gauge physics can be regularized
 by regularizing its quasi-Lie algebras.
 This eliminates gauge groups as well as theory singularities.
This will be taken up elsewhere.

\section {Higher-order quantum set theory}
\label{S:QST}
Classical set theory iterates the power-set functor to form the 
space of all ``regular" (ancestrally finite, hereditariliy finite) sets.
A regular set theory might therefore iterate the Fermi quantification functor \cite{FINKELSTEIN1996}, as follows.

The Peano $\iota$, with
\BEQ
\{a,b,c,\dots\}:=\{a\}\{b\}\{c\}\dots=\iota a \;\iota b \;\iota c \;\dots,
\EEQ
defines membership $a\in b$:
\BEQ
a \in b\quad :\equiv \quad \iota a \subset b
\EEQ
 
Let $\1S$ designate the classical algebra of finite sets
finitely  generated from the empty set 1
by bracing $\iota x=\{x\}$ and the disjoint union $x \vee y$
(a group product with identity 1, the empty set).
Sets of $\1S$ are here called {\em perfinite} 
(elsewhere, ancestrally or hereditarily finite).
They are finite, and so are their elements, and their elements, and so forth,
all the way down to the empty set.
Let $\bigvee s$ be the  set of finite subsets of $s$. Then
\BEQ
\bigvee :\1S\to \1S = \bigvee \1S.
\EEQ
An element of $\1S$ is a set or simplex whose vertices may be sets or simplices.
$\1S$ is supposedly complex enough to represent any finite classical structure.

A quantum analogue  $\8{\1S}$ is a kind of linearization of $\1S$:

For any quadratic space $S$,
let  $\bigsqcup S$ designate the Clifford algebra 
of finite-degree polynomials
over $S$, modulo the  exclusion principle
\BEQ
\forall s\in S: s\sqcup s=0\/.
\EEQ
$\bigsqcup S$ and $\sqcup$  correspond to the classical power set
and the symmetric union ({\sc xor}).
 If  $\3P_1$ is a one-fermion probability space 
 then $\3P=\Cliff \3P_1$ is the many-fermion probability algebra.

Each quantum subclass of a 
system is associated with a subspace
$\3C\subset \3P$ in the probability space of the system, and so
with
a Clifford probability vector $e_{\3C}$,
a  top vector  of the Clifford algebra $\Cliff \3C\subset \Cliff \3P$.

Then define $\iota:\3P\to \Cliff\3P$ as a Cantor brace, modulo linearity:
\BEQ
  \forall p\in \3P: \iota p:=\{p\}, \quad
   \mbox{mod}\quad \iota(ax+by)\equiv a\,\iota x + b\, \iota y\/.
\EEQ

Take $\8{\1S}$ (as a first trial) to be the least Clifford algebra that is its own Clifford algebra:
\BEQ
\bigsqcup :\8{\1S}\to \bigsqcup \8{\1S}= \8{\1S}.
\EEQ
Call the quantum structures with  probability vectors in $\8{\1S}$  {\em quantum sets}\/.
The quantum set is supposedly complex enough to represent any finite quantum  structure.

Table 1 arranges  basic probability vectors $1_n$ of $\8{\1S} $  by rank  $r$ and serial number $n$.
\BTA [h]
\caption{Quantum and classical sets $1_n$ by rank $r$ and serial number $n$}
{\boldmath
\[
\begin{tabular}{|c|ccccccccccccc|} 
\hline
{\large  6} \vrule height 30pt depth 0pt width 0pt 

&
\huge$\stackrel{\6{\6{\6{\6{\6{\6{\Z}}}}}}}{}$&\huge$
\stackrel {\6{\6{\6{\6{\6{\6{\Z}}}}}}\,\6{\Z}}{}$&\huge$
\stackrel {\6{\6{\6{\6{\6{\6{\Z}}}}}}\,\6{\6{\Z}}}{}$&\huge$
 \stackrel {\6{\6{\6{\6{\6{\6{\Z}}}}}}\,\6{\6{\Z}}\,\6{\Z}}{}$&\huge$
\stackrel {\6{\6{\6{\6{\6{\6{\Z}}}}}}\,\6{\6{\6{\Z}}}}{}$&\huge$
\stackrel {\6{\6{\6{\6{\6{\6{\Z}}}}}}\,\6{\6{\6{\Z}}}\,\6{\Z}}{}$&\huge$
\stackrel {\6{\6{\6{\6{\6{\6{\Z}}}}}}\,\6{\6{\6{\Z}}}\,\6{\6{\Z}}}{}$&\huge$
\stackrel {\6{\6{\6{\6{\6{\6{\Z}}}}}}\,\6{\6{\6{\Z}}}\,\6{\6{\Z}}\,\6{\Z}}{}$&\huge$
\stackrel {\6{\6{\6{\6{\6{\6{\Z}}}}}}\,\6{\6{\6{\Z}}\,\6{\Z}}}{}$&\huge$
\stackrel {\6{\6{\6{\6{\6{\6{\Z}}}}}}\,\6{\6{\6{\Z}}\,\6{\Z}}\,\6{\Z}}{}$&\huge$
\stackrel{\6{\6{\6{\6{\6{\6{\Z}}}}}}\,\6{\6{\6{\Z}}\,\6{\Z}}\,\6{\6{\Z}}}{}$&\huge$
\stackrel{\6{\6{\6{\6{\6{\6{\Z}}}}}}\,\6{\6{\6{\Z}}\,\6{\Z}}\,\6{\6{\Z}}\,\6{\Z}}{}$&\large$^
 {\dots}
 $
 \cr
  \vrule height 12pt depth 0pt width 0pt&$\!^{\hexp 6}$&\large $^{\dots}$&\large $^{\dots}$&\large $^{\dots}$&&&&&&&&&\cr
\hline
{\large  5}&
\vrule height 30pt depth 0pt width 0pt
\huge$
\stackrel {\6{\6{\6{\6{\6{\Z}}}}}}{}$&\huge$
\stackrel{\6{\6{\6{\6{\6{\Z}}}}}\,\6{\Z}}{}$&\huge$
\stackrel{\6{\6{\6{\6{\6{\Z}}}}}\,\6{\6{\Z}}}{}$&\huge$
\stackrel{\6{\6{\6{\6{\6{\Z}}}}}\,\6{\6{\Z}}\,\6{\Z}}{}$&\huge$
\stackrel{\6{\6{\6{\6{\6{\Z}}}}}\,\6{\6{\6{\Z}}}}{}$&\huge$
\stackrel{\6{\6{\6{\6{\6{\Z}}}}}\,\6{\6{\6{\Z}}}\,\6{\Z}}{}$&\huge$
\stackrel{\6{\6{\6{\6{\6{\Z}}}}}\,\6{\6{\6{\Z}}}\,\6{\6{\Z}}}{}$&\huge$
\stackrel {\6{\6{\6{\6{\6{\Z}}}}}\,\6{\6{\6{\Z}}}\,\6{\6{\Z}}\,\6{\Z}}{}$&\huge$
\stackrel {\6{\6{\6{\6{\6{\Z}}}}}\,\6{\6{\6{\Z}}\,\6{\Z}}}{}$&\huge$
\stackrel{\6{\6{\6{\6{\6{\Z}}}}}\,\6{\6{\6{\Z}}\,\6{\Z}}\,\6{\Z}}{}$&\huge$
\stackrel{\6{\6{\6{\6{\6{\Z}}}}}\,\6{\6{\6{\Z}}\,\6{\Z}}\,\6{\6{\Z}}}{}$&\huge$
\stackrel{\6{\6{\6{\6{\6{\Z}}}}}\,\6{\6{\6{\Z}}\,\6{\Z}}\,\6{\6{\Z}}\,\6{\Z}}{}$&\large $
^{\dots}
 $\cr
& $\!^{\hexp{5}}$&\large $^{\dots}$&\large $^{\dots}$&\large $^{\dots}$&&&&&&&&&\cr
\hline
{\large   4}& \vrule height 30pt depth 0pt width 0pt
\huge $
\stackrel {\6{\6{\6{\6{\Z}}}}}{}$&\huge$   
\stackrel{\6{\6{\6{\6{\Z}}}}\,{\6{\Z}}}{}$&\huge$
\stackrel{\6{\6{\6{\6{\Z}}}}\,{\6{\6{\Z}}}}{}$&\huge$
\stackrel{\6{\6{\6{\6{\Z}}}}\,{\6{\6{\Z}}\,\6{\Z}}}{}$&\huge$
\stackrel{\6{\6{\6{\6{\Z}}}}\,{\6{\6{\6{\Z}}}}}{}$&\huge$
\stackrel{\6{\6{\6{\6{\Z}}}}\,{\6{\6{\6{\Z}}}}\,\6{\Z}}{}$&\huge$
\stackrel{\6{\6{\6{\6{\Z}}}}\,{\6{\6{\6{\Z}}}\,\6{\6{\Z}}}}{}$&\huge$
\stackrel {\,\6{\6{\6{\6{\Z}}}}\,{\6{\6{\6{\Z}}}\,\6{\6{\Z}}}\,\6{\Z}}{}$&\huge$
\stackrel {\6{\6{\6{\6{\Z}}}}\,{\6{\6{\6{\Z}}\,\6{\Z}}}}{}$&\huge$
\stackrel {\6{\6{\6{\6{\Z}}}}\,{\6{\6{\6{\Z}}\,\6{\Z}}}\,\6{\Z}}{}$&\huge$
\stackrel {\6{\6{\6{\6{\Z}}}}\,{\6{\6{\6{\Z}}\,\6{\Z}}}\,\6{\6{\Z}}}{}$&\huge$
\stackrel {\6{\6{\6{\6{\Z}}}}\,{\6{\6{\6{\Z}}\,\6{\Z}}}\,\6{\6{\Z}}\,\6{\Z}}{}$&\large 
$^{\dots}$\cr
 &16 &17&18&19&20&21&22&23&24&25&26&27&${{\dots}}$\cr
\hline 
{\large   3} & \vrule height 30pt depth 0pt width 0pt \huge$
\stackrel {\6{\6{\6{\Z}}}}{}\kern0pt$&\huge$  
\stackrel{{\6{\6{\6{\Z}}}\,\6{\Z}}}{}\kern3pt$&\huge$ 
\stackrel{\6{\6{\6{\Z}}}\,\6{\6{\Z}}}{}\kern3pt$&\huge$
\stackrel{{ \6{\6{\6{\Z}}} \, \6{\6{\Z}} \, \6{\Z}}}{}\kern3pt$&\huge$
\stackrel{{\6{\6{\6{\Z}}\,\6{\Z}}}}{}\kern3pt$&\huge$
\stackrel{{\6{\6{\6{\Z}}\,\6{\Z}}}\,{\6{\Z}}}{}\kern3pt$&\huge$ 
\stackrel{{\6{\6{\6{\Z}}\,\6{\Z}}}\,{\6{\6{\Z}}}}{}\kern3pt$&\huge$
\stackrel{{\6{\6{\6{\Z}}\,\6{\Z}}}\,{\6{\6{\Z}}}\,\6{\Z}}{}\kern3pt$&\huge$
\stackrel{{\6{\6{\6{\Z}}\,\6{\Z}}}\,{\6{\6{\6{\Z}}}}}{}\kern3pt$&\huge$
\stackrel{{\6{\6{\6{\Z}}\,\6{\Z}}}\,{\6{\6{\6{\Z}}}}\,\6{\Z}}{}\kern3pt$&\huge$
\stackrel {{\6{\6{\6{\Z}}\,\6{\Z}}}\,{\6{\6{\6{\Z}}}}\,\6{\6{\Z}}}{}\kern3pt$&\huge$ 
\stackrel{{\6{\6{\6{\Z}}\,\6{\Z}}}\,{\6{\6{\6{\Z}}}}\,\6{\6{\Z}}\,\6{\Z}}{}$&\cr
 &4 &5&6&7&8&9&10&11&12&13&14&15&\cr
\hline
{\large   2}  & \vrule height 30pt depth 0pt width 0pt 
\huge $
\stackrel {\6{\6{\Z}}}{}$&\huge$
\stackrel{{\6{\6{\Z}}}\,\6{\Z}}{}$&&&&&&&&&&&\cr
&2&3&&&&&&&&&&&\cr
 \hline
{\large    1} & \vrule height 20pt depth 10pt width 0pt
\Huge $\stackrel {\6{\Z}}{}$&${}$&&&&&&&&&&&\cr
&1 &&&&&&&&&&&&\cr
\hline
{\large   0}  & \vrule height 20pt depth 0pt width 0pt
\kern1pt
{\Z}
 \vrule height 10pt depth 0pt width 0pt
 &&&&&&&&&&&&\cr
&\kern0pt  0 &&&&&&&&&&&&\cr
\hline
\hline
{\large \em r}  &\vrule height 20pt depth 0pt width 0pt&&&&&&{\large${1_n}$}&&&&&&\cr
&&&&&&&{\large \em n}\vrule height 20pt depth 0pt width 0pt&&&&&&\cr
\hline
\end{tabular}
\]
}
\ETA

\subsection{Spin structure of $\8{\1S} $}
\label{S:SPIN}

For each rank $r$,  $\8{\1S} [r]$ is naturally a spinor space:
\begin{itemize}

\item $D[r]=\hexp r$ is its dimension.

\item $\8{\1S} [r-1]$ is its Cartan semivector space.

\item $\3W[r-1]:=  \8{\1S} [r-1]\oplus \Dual\; \8{\1S} [r-1]$ is its underlying quadratic space.

\item $\SO(D[r-1],D[r-1])$ is its orthogonal group. 

\item There is a neutral symmetric Pauli form $\beta[r]: \8{\1S} [r]\to \Dual \8{\1S} [r]$
for which the first grade $\gamma^w\in \Cliff[r]$ are hermitian symmetric.

\item The Pauli form can be chosen to be a Berezin integral with respect to the top 
Grassmann element (or volume element) $\gamma^{\top }\in \3W[r]$:
\BEA
\beta[r-1]:\3W[r-1]\ox\3W[r-1]&\to &\1R,\cr
\forall\psi=w\oplus w'\in \3W[r]:\quad  \|\psi\|_{r-1}= \beta[r] \psi \psi&:=&\int d\gamma^{\top}\psi^2\/ =\partial_{(\gamma^{\top})}\psi^2.
\EEA

\item $\Cliff( \3W[r-1]) \cong \Endo_{\0{Vec}}\8{\1S} [r]$:
the algebra of linear operators on the spinor space is isomorphic as algebra to the associated Clifford algebra $\Cliff[r]$.

\end{itemize}
This $\beta$ is just the $\beta$ of Pauli  and Chevalley expressed
in the more powerful notation used by physicists.
Since $L^2(\3M)$ designates a quadratic space defined by a quadratic Lebesgue integral
over $\3M$, 
write  the quantum space defined by a quadratic Berezin integral  over $\3W$ as  $B^2(\3W)$.

Every real Grassmann algebra $\3G= \Grass {N\1R}$ is a spinor space
for the orthogonal group whose quadratic space
$\3W$  is the direct sum of the polar and axial vectors of $\3G$,
grades 1 and $N-1$ of $\3G$:
\BEQ
\3W=\Grade_1 \3G\oplus \Grade_{N-1} \3G.
\EEQ
The  norm on $\3W$ is the quadratic Berezin form 
\BEA
\beta:\3W\ox\3W&\to &\1R,\cr
\forall\psi=w\oplus w'\in \3W:\quad  \|\psi\|= \beta\psi \psi&:=&\int d\gamma^{\top}\psi^2\/ =\partial_{(\gamma^{\top})}\psi^2.
\EEA

This imbedding of the quadratic space $\3W$  in its 
spinor space is isometric
but not invariant under  $\spin(\3W)$,
which mixes  $\Grade_1 G$ and $ \Grade_{N-1} G$.

\section{Revised quantum set theory}
\label{S:ANTICANTOR}
Here are some adaptations of $\8{\1S}$ to current physics.

\subsection{Bosons}
\label{S:BOSONS}
The number of times one set  belongs to another ($a\in b$)
is either 0 or 1. 
In this respect classical sets have Fermi (odd) statistics. 
Classical thought did not allow for Bose (even) statistics, 
which  grossly violates the Leibniz doctrine 
that indistinguishable objects are one.
Nor does $\8{\1S} $ describe elementary bosons.
The Standard Model, however, requires them.
Moreover,  ${a}$ and $b$ are monads (first-grade
elements) of the  space $\8{\1S} $, 
hence fermionic, then $\{a,b\}$ is  a fermionic monad too, 
although $a\v b$ is an approximate boson and an exact palevon.
Two odds make an odd in set theory,
and  an even in nature.
$\8{\1S} $  violates conservation of statistics. 

This is resolved by modeling the even quanta
as pairs of odd quanta.
These pairs can then be 
associated by unition.
To be sure the result is odd, a unit set.
But every rank $r$ in $\8{\1S}$ has its own product 
$\stackrel{r}{\v}$,
and this can be applied to  rank $r+1$ as well as $r$:
One disunites,  multiplies, and reunites.
The spin-statistics relation refers to one rank in the theory.

\subsection{Reducibility}
\label{S:REDUCIBILITY}
The power set contains subsets of every cardinality up to the maximum.
In the quantum theory the 
corresponding construction leads to a  representation
of the group of the previous rank
that is reducible by degree.
The known quanta provide irreducible representations,
and interact by irreducible couplings.

Nevertheless the Standard Model  uses the unreduced classical brace
$\{\dots\}$ 
to assemble (say) a fermion probability vector from 
orbital, isospin, spin, and other probability vectors.
The same practice works in the regular theory $\6{\1S}$.

\subsection{Irreversibility}
\label{S:IRREVERSIBILITY}
$\8{\1S} $ is founded on the highly asymmetric operation $\iota$
of degree ${1} \choose{n}$.
All the interactions in the Standard Model are symmetric with respect to the exchange of input and output.
If the input has degree 2, then so does the output,
and the tensor is of degree  $2 \choose {2}$.
This degree does not occur among the basic operations in $\8{\1S}$.
But it is a term in $\iota^{\beta}\iota$, 
where $\beta$ is the Pauli adjoint operation.

{\em Proposition}: $\fr d_r$ is asymptotically neutral:
\BEQ
\lim_{n\to \infty}\; \frac {p_r}{n_r}\to 1\/.
\EEQ
This approaches the canonical signature of the probability space. 
The proof is straightfoward.
For finite ranks $r>2$, however, the space is never exactly neutral.
One must wonder  whether this slight departure from symmetry between
input and outtake can be used to represent
a slight asymmetry between matter and anti-matter,
or a non-zero vacuum energy density.

Naturally Cartan based his spinor theory
on a classical space-time.
There is none in nature, so
the generators of the spin group should be  interpreted 
in the earlier manner of Schur \cite{SCHUR1911},
as binary flips, not rotations, 
and then quantized.
The Clifford algebra $\Cliff(n)$ now represents a finite quantum group
corresponding to the classical finite group $2^n$.

The probability vector space of the Standard Model is a tensor product
of fermionic Clifford algebras  and Bose symmetric tensor algebras.
Regularization replaces every Bose Lie algebra by a Palev algebra.
But the fermion algebra already contains many palevon ones.
This leads one to suspect that the bosons of the regularized Standard Model can be economically represented as fermion pairs
held together by binding rather than by unition.

Such a possibility was already raised by the 
de Broglie two-neutrino  photon,  
and  the four-neutrino  graviton considered 
and rejected by Feynman.
The main obstacle to such constituent theories is that according to the Heisenberg indeterminacy principle, fermions near each other
in position must be far apart in momentum. 
Then  they require a correspondingly high interaction-energy for their binding.
But experiment finds no such intense interaction
but only the asymptotic freedom implied by the Standard Model.

In a Feynman or Yang quantum space-time, however, 
the operator $\Qi\hbar$ that replaces $i\hbar$ has a finite spectrum of magnitudes with extreme values $\pm \hbar$.
Presumably a self-organization akin to ferromagnetization
freezes $\Qi\hbar$ to its maximum value $i\hbar$.
The Heisenberg uncertainty principle is then weakened wherever  a local disorganization reduces the magnitude of
$\Qi \hbar$.
Such a local defect might allow two leptons to bind into a photon, say.
This re-opens the question of a di-fermion theory of 
the gauge bosons. 

\section{Gratitude}
 
To Roger Penrose and  Richard Feynman for kindly
 showing me  their seminal spin quantizations of space and space-time
 before publication.
To  O. B. Bassler,
James {Baugh}, Walter L. Bloom, Jr., Dustin Burns, David Edwards,
 {Shlomit} Ritz Finkelstein, Andrei {Galiautdinov}, Dennis Marks,
  Zbigniew Oziewicz, Tchavdar Dimitrov Palev,
   Heinrich {Saller}, Stephen Selesnick,
  Abraham Sternlieb,  Sarang Shah,
and Frank (Tony) Smith,
 for helpful and enjoyable discussions.
 
 To Lynn Margulis for introducing me to post-Darwinian symbiogenetic
 evolution in  her Lindisfarne lectures.
 
To Cecylia Arszewski for leading me to Unger's 
most helpful radical pragmatism.
 
To Prof. V. K. Dobrev for inviting me
to this retreat of the Bulgarian Academy of Science.


\begin{thebibliography}{00}
 
 \bibitem{BOPP1950}
 F. Bopp and R. Haag.
 \"Uber die M\"oglichkeit von Spinmodellen.
 {\em Zeitschrift f\"ur  Naturforschung} 5a:644 
(1950).


\bibitem{DIRAC1974}
P. A.M. Dirac.
{\em Spinors in Hilbert Space}\/.
Plenum, New York (1974).


\bibitem{FEYNMAN1941}
R. P. Feynman.
Personal communication ca. 1961.
Feynman did this in about 1941,
before his work on the Lamb shift,
and probably published
this formula in a footnote,
but we did not find the reference.

\bibitem{FINKELSTEIN1996} Finkelstein, D. 
{\it Quantum Relativity}.
Heidelberg: Springer, 1996.

\bibitem{HALDANE1991}
F. D. M. Haldane.
``Fractional Statistics" in Arbitrary Dimensions: A Generalization of the Pauli Principle. , Physical Review  Letters  67: 937 ~1991.

\bibitem{INONU1952} E. In\"on\"u and E. P. Wigner.
On the contraction of groups and their representations.
{\it Proceedings of the National Academy of Sciences} 
39:510-525 (1952).

\bibitem{OPERA2011}
OPERA Collaboration.
Measurement of the neutrino velocity with the OPERA detector in the CNGS beam. arXiv:1109.4897   (2011)
 
 
\bibitem{PALEV1977}
T. D. Palev.
Lie algebraical aspects of the quantum statistics. {U}nitary
quantization ({A}-quantization).
Joint Institute for Nuclear Research Preprint JINR E17-10550.
Dubna (1977).
hep-th/9705032.

\bibitem{PALEV2002}
T. D. Palev
and
J. Van der Jeugt.
Jacobson generators, Fock representations and statistics
of $sl(n+1)$.
{\em Journal of Mathematical Physics} 43:3850-3873 (2002).

T.D. Palev and J. Van der Jeugt,
Jacobson generators, Fock representations and statistics of sl(n+1).
J. Math. Phys. 43 (2002), . 
 
\bibitem{PENROSE1971}
R. Penrose.
Angular momentum: an approach to combinatorial
space-time.
In T. Bastin (ed.),
{\em Quantum Theory and Beyond}\/,
151--180,
Cambridge 1971.
Penrose kindly shared much of this seminal work
with me ca. 1960.

\bibitem{SALLER2006a}
H. Saller.
{\em Operational Quantum Theory I. Nonrelativistic Structures.}
Springer, New York (2006).


\bibitem{SCHUR1911}
I. Schur.
{\"U}ber die {D}arstellung der symmetrischen und der alternierenden 
{G}ruppe durch gebrochene lineare {S}ubstitutionen.
{\em Journal f\"ur die reine und angewandte {M}athematik}
139:155--250 (1911)

 \bibitem{SEGAL1951}
I. E. Segal.
A class of operator algebras which are determined by groups.
Duke Mathematical Journal
18:221--265 (1951).
Especially {\S}6A.

\bibitem{UNGER2007} 
R. M. Unger.  {\em The Self Awakened:  Pragmatism Unbound}\/.
Harvard University Press (2007).
 
 \bibitem{YANG1947}
C. N. Yang.
On Quantized Space-Time.
{\em Physical Review}  72:874
(1947).


 \end{thebibliography}
\end{document}